\documentclass[a4paper,11pt]{article}

\usepackage{amsmath}
\usepackage{latexsym}
\usepackage{amsfonts}
\usepackage{amssymb}
\usepackage{graphicx}
\usepackage{txfonts}
\usepackage{wasysym}
\usepackage{enumitem}
\usepackage{adjustbox}
\usepackage{ragged2e}
\usepackage{tabularx}
\usepackage{changepage}
\usepackage{setspace}
\usepackage{hhline}
\usepackage{multicol}
\usepackage{float}
\usepackage{multirow}
\usepackage{makecell}
\usepackage{fancyhdr}
\usepackage[toc,page]{appendix}
\usepackage[utf8]{inputenc}
\usepackage[T1]{fontenc}
\usepackage{hyperref}

\hypersetup{
colorlinks=true,
linkcolor=blue,
filecolor=magenta,
urlcolor=cyan,
}
\usepackage[a4paper,left=2cm,right=2cm,bottom=2cm,top=2cm]{geometry}
\everymath{\displaystyle}

\begin{document}
\sloppy
\begin{center}\textbf{{\fontsize{14pt}{14pt}\selectfont Software for full-color 3D reconstruction of the biological tissues internal structure \\}}\end{center} \par
\noindent
\begin{center}Khoperskov A.V.\textsuperscript{1[0000-0003-0149-7947]}, Kovalev M.E.\textsuperscript{1[0000-0001-5136-4048]}, Astakhov A.S.\textsuperscript{1[0000-0002-9905-568X]}, Novochadov V.V.\textsuperscript{2[0000-0001-6317-7418]}, Terpilovskiy A.A.\textsuperscript{3[0000-0001-8200-044X]}, Tiras K.P.\textsuperscript{4[0000-0002-1853-8285]}, Malanin D.A.\textsuperscript{5[0000-0002-8065-6506]}\end{center} \par
\begin{center}
{\fontsize{9pt}{3pt}\selectfont \textsuperscript{1} Institute of Mathematics and Information Technologies, Volgograd State University, Volgograd, Russia} \vspace{6pt}
{\par\fontsize{9pt}{3pt}\selectfont \textsuperscript{2} Institute of Natural Sciences, Volgograd State University, Volgograd, Russia\par
khoperskov@volsu.ru, a.s.astahov@volsu.ru, biobio@volsu.ru} \par
{\fontsize{9pt}{3pt}\selectfont \textsuperscript{3} The laboratory of virtual biology, Ltd., Moscow, Russia}\par
{\fontsize{9pt}{3pt}\selectfont \textsuperscript{4} Institute of Theoretical and Experimental Biophysics of RAS, Moscow, Russia} \par
{\fontsize{9pt}{9pt}\selectfont \textsuperscript{5} Volgograd State Medical University, Volgograd, Russia}\end{center} \par
{\fontsize{9pt}{9pt}\selectfont \textbf{Abstract. }A software for processing sets of full-color images of biological tissue histological sections is developed. We used histological sections obtained by the method of high-precision layer-by-layer grinding of frozen biological tissues. The software allows restoring the image of the tissue for an arbitrary cross-section of the tissue sample. Thus, our method is designed to create a full-color 3D reconstruction of the biological tissue structure. The resolution of 3D reconstruction is determined by the quality of the initial histological sections. The newly developed technology available to us provides a resolution of up to 5 - 10  $  \mu  $m in three dimensions.} \par
\noindent
\begin{flushleft}\textbf{Keywords:} Scientific software, Computer graphics, 3D reconstruction, Biological tissues, Image processing.\end{flushleft} \par  \vspace{6pt}
\noindent
\centerline{\bf { Introduction }}
\noindent

Different approaches and methods of 3D-visualization of biological tissues are discussed depending on the scientific goals and objectives of practical applications [\ref{b1}{1}, \ref{b2}{2}, \ref{b14}{14}]. A significant number of software complexes, libraries and specialized systems of scientific and technological visualization for three-dimensional digital biomedicine have been developed [\ref{b7}{7}, \ref{b9}{9}, \ref{b12}{12}, \ref{b24}{24}, \ref{b25}{25}]. Approaches using medical X-ray tomography are very effective for 3D-reconstruction of tissues/organs [\ref{b8}{8}]. We can also highlight some areas that are associated with the localization of implanted biomaterials [\ref{b8}{8}, \ref{b26}{26}], the colonoscopy, the ultrasonic sounding [\ref{b9}{9}], the stereoscopic fluorescence imaging, the multispectral magnetic resonance image analysis [\ref{b16}{16}], the single photon emission computed tomography (CT) [\ref{b11}{11}], the electron tomography [\ref{b6}{6}], the use of combined methods [\ref{b9}{9}]. The transition begins from micro-CT to nano-CT [\ref{b8}{8}, \ref{b20}{20}]. ~  \par
\vspace{6pt}
This work is addressed to the anatomical or destructive tomography (biotomy) approach [\ref{b17}{17}, \ref{b23}{23}]. A newly developed method is based on making of a set of high-quality images of biological tissue histological sections using the high-precision grinding of a pre-frozen biological sample [\ref{b22}{22}]. \par
\vspace{6pt}
The advantages of the proposed approach: \par
\begin{enumerate}
\item\noindent
A high-quality photography offering all the benefits of the raster graphics. \par
\item\noindent
A realistic color rendering. \par
\item\noindent
Very high accuracy up to several microns. \par
\item\noindent
Absence of "screening interference"\ in contrast to non-destructive methods [\ref{b26}{26}]. \par
\item\noindent
High speed access to information about individual fragments of the tissue. \par
\end{enumerate}
The main disadvantage of the proposed method is principle impossibility of living tissue usage due to its physical destruction during the process of preparation of a set of images slices. Another disadvantage of the method associates with utilization of digital raster photographs. 3D modeling can improve the efficiency of various solving aims tasks, including endoprosthetics [\ref{b5}{5}, \ref{b8}{8}, \ref{b10}{10}, \ref{b15}{15}, \ref{b26}{26}, \ref{b27}{27}]. \par
\vspace{6pt}
\centerline{\textbf {Making of an images set of histological sections}}


\begin{figure}[H]
\begin{center}
\includegraphics[width=4.79in,height=2.37in]{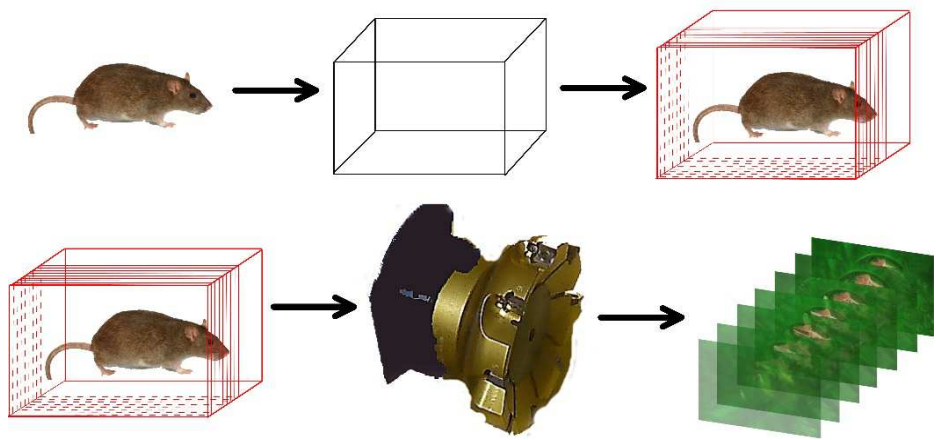}
\caption{The sequence of steps for making of an images set of biological tissue histological sections}
\end{center}
\label{fig1}
\end{figure}


\noindent

The process of producing of images set includes the following steps (Fig. \ref{fig1}{1}): \par
\begin{enumerate}
\item\noindent Selection of a biological object in accordance with the objectives of the study. \par
\item\noindent
A formation of the original sample by fixing a biological object by a casting material in a three-dimensional mold. \par
\item\noindent
The original sample of the biological object is rapidly frozen in the form of a rectangular parallelepiped. The fixation of the biological object is ensured by at cryogenic temperatures (T  $  \leq -72^{\circ}  $C), after that it is filled by the casting material (at T  $  \leq  -25^{\circ}  $C) [\ref{b19}{19}].
\item\noindent
The sample is placed into a special device for layer-by-layer grinding. \par
\item\noindent
An optical system adjustment is required due to the sample thickness change as a result of layer-by-layer removing. \par
\item\noindent
Visual analysis of the images allows distinguishing photos with a blurred focus, highly modified illumination and other optical artifacts (removing images of poor quality if it is necessary). \par
\end{enumerate}
\par As a result, we have an ordered snapshot sequence of biological tissue slices with step \textit{ $  \Delta  $z}, whose resolution determined by sample size and used object-glass. The \textit{ $  \Delta  $x} and \textit{ $  \Delta  $y} resolution of the equipment available at The Laboratory of Virtual Biology, Ltd is 6-20  $  \mu  $m, while the \textit{ $  \Delta  $z} resolution is 5 - 20  $  \mu  $m.
 \vspace{12pt}
\par
\centerline{\bf {Creation of $\Theta$-sections and software development}}

\begin{figure}[H]
\begin{center}
\includegraphics[width=3.73in,height=3.1in]{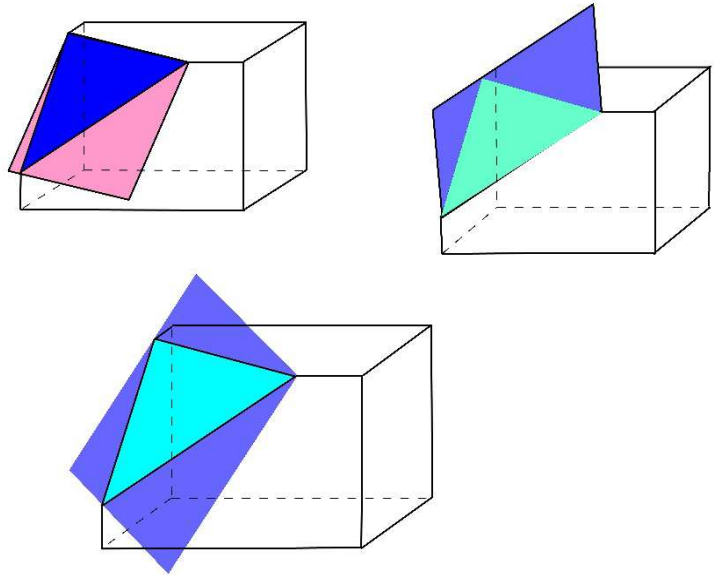}
\caption{The examples of the $\Theta$-rectangles in a triangular section}
\end{center}
\label{fig2}
\end{figure}

Let us introduce a $\Theta$-rectangle ("theta-rectangle") to denote a rectangle inscribed into the polygon of the sample section. Various ways of $\Theta$-rectangle specification in a triangular section are shown in Fig.\ref{fig2}{2}. \par


Hereafter the RGB color system is applied. Each element of the mapping color table \textit{P} is considered like a vector in the three-dimensional space. Thereby, within the closure region of the sample space the vector field $  \vec{\Omega}  \left( x,y,z \right) = \{ R,G,B \}  $ is a set that converts it into a color space. To construct an image on the sample slice plane, the -rectangle of this section have to be divided into pixels. Each pixel is associated with a color depending on its spatial position in a vector field $ \vec{\Omega}  $. \par
A set of slices images is the initial data
 \begin{equation}\label{eq1}
   z_{k}= \left( k-1 \right) \Delta z,
 \end{equation}

where the \textit{k}-th image number (\textit{k=1,…,M}) is determined by the \textit{z}-coordinate of the histological section. \par
 \hspace*{0.16in} The sample vertex in the upper left corner of its frontal face relatively to the camera location during the record is taken as origin. Let us Consider a numbered set of raster images (\ref{eq1}) of size
 $P_{x} \times P_{y}$, where $P_{x}$ and $P_{x}$ are the number\textbf{s} of pixels for the two corresponding sides. Such an initial set of images is also called a digital set of base slices. \par
\noindent
\par To reconstruct an object in the form of a three-dimensional model, a method for color interpolation in the space between each adjoining pair of images in the plains $\textit{z}_{\textit{k}}$ and $\textit{z}_{\textit{k+1}}$ should be specified. The entire set of initial images is considered as a three-dimensional matrix $ \widehat{P} = \left( i,j,k \right)  $
, whose elements are individual pixels of images from the set. Their addressing is determined by the number of the \textit{k}-th layer (a sequence number of a photo), the number of the \textit{i}-th row and the number of the \textit{j}-th column in the matrix of pixels of the \textit{k}-th photo. We have used discrete functions defined on a three-dimensional grid $\textit{(x}_{\textit{i}}\textit{,y}_{\textit{j}}\textit{,z}_{\textit{k}}\textit{)}$.  \par
\noindent
\par The statement that the vector is orthogonal to the normal vector to the plane described by the equation $\textit{Ax + By + Cz + D = }0$ is equivalent tantamount to the following expression: \par
\begin{equation}\label{eq2}
A \xi _{x}+C \xi _{z}=0 .
\end{equation}
\par The vector \textit{ $  \vec{\eta}  $} obeys the conditions of orthogonality to the vector \textit{ $  \xi  $} and the normal to the section plane. The latter means that its direction may be computed as the vector product of the vector \textit{ $  \xi  $}  by the normal vector to the section plane: \par
\begin{equation}\label{eq3}
 \vec{\eta} = \left[  \xi ,n \right] =det
 \begin{pmatrix}
 e_{x} ~ e_{y} ~ e_{z}\\
 \xi_{x} ~ \xi_{y} ~ \xi_{z}\\
 A ~ B ~ C
 \end{pmatrix} ,
\end{equation}
where \textit{n} is the normal vector to the plane of the section, $e_{x}$, $e_{y}$, $e_{z}$ are the unit vectors along the positive directions of the \textit{x}-, \textit{y}- and \textit{z}-axes respectively in the reference system associated with the sample. \par
Finally, taking into account the fact that the vectors $\xi$ and $\vec{\eta}$ have a unit length, we obtain two more equations: \par
\begin{equation}\label{eq4}
\sqrt[]{ \xi _{x}^{2}+~ \xi _{z}^{2}}=1,
\end{equation}
\begin{equation}\label{eq5}
\sqrt[]{ \vec{\eta}_{x}^{2}+ \vec{\eta} _{y}^{2}+ \vec{\eta} _{z}^{2}}=1.
\end{equation}
\par From equations (\ref{eq2}-\ref{eq5}), we can calculate the coordinate values for the vectors \textit{ $  \xi  $} and \textit{ $  \vec{\eta}  $}: \par
\begin{equation}\label{eq6}
\xi =~ \left( \frac{C}{\sqrt[]{A^{2}+C^{2}}},~0,-\frac{A}{A^{2}+C^{2}} \right),
\end{equation}
\begin{equation}\label{eq7}
\vec{\eta} =~ \left( -\frac{AB}{L},~\frac{A^{2}+C^{2}}{L},-\frac{BC}{L} \right),
\end{equation}
\begin{equation}\label{eq8}
L=\sqrt[]{A^{2}B^{2}+ \left( A^{2}+C^{2} \right) ^{2}+B^{2}C^{2}} .
\end{equation}
To solve the problem of the long time required a data reading from the hard disk, we choose a special basis on the plane of the section. We denote by \textit{p}, \textit{q} the unit vectors along the directions of the abscissa and ordinate axes, respectively, in the reference system associated with the section plane. Thus, the vector \textit{p} is orthogonal to the unit vector directed along the applicate axis in the reference system associated with the sample: \par
\begin{equation}\label{eq9}
p_{z}=0,
\end{equation}
where $p_{z}$ is the projection of the vector $p$ onto the given axis. \par
 \hspace*{0.16in} Moreover, it is required adding equations denoting the orthogonality of the vector \textit{p} to the normal vector to the plane of the section, and the orthogonality of the vector \textit{q} to both of them, as well as the condition that the length of the vectors \textit{p} and \textit{q} is equal to one. Accounting for all the listed above conditions we obtain the following equations: \par
\begin{equation}\label{eq10}
Ap_{x}+Bp_{y}=0,
\end{equation}
\begin{equation}\label{eq11}
q=det
 \begin{pmatrix}
 e_{x} ~ e_{y} ~ e_{z}\\
 p_{x} ~ p_{y} ~ p_{z}\\
 A ~~ B ~~ C
 \end{pmatrix}
 ,
\end{equation}
\begin{equation}\label{eq12}
\sqrt[]{p_{x}^{2}+p_{y}^{2}=1},
\end{equation}
\begin{equation}\label{eq13}
\sqrt[]{q_{x}^{2}+q_{y}^{2}+q_{z}^{2}=1} .
\end{equation}
From the equations (\ref{eq9}-\ref{eq13}), we can determine the values of the coordinates of the basis vectors in the reference system associated with the sample: \par
\begin{equation}\label{eq14}
p= \left( -\frac{B}{\sqrt[]{A^{2}+B^{2}}},~\frac{A}{\sqrt[]{A^{2}+B^{2}}},0 \right),
\end{equation}
\begin{equation}\label{eq15}
q= \left( -\frac{AC}{M},-~\frac{BC}{M},\frac{A^{2}+B^{2}}{M} \right),
\end{equation}
\noindent
\begin{equation}\label{eq16}
M=\sqrt[]{A^{2}C^{2}+B^{2}C^{2}+ \left( A^{2}+B^{2} \right) } .
\end{equation}
The equations (\ref{eq14}) and (\ref{eq15}) specify the directions of the basis vectors of such reference system, associated with the cross-section plane, for which the natural order of pixel bypass would guarantee the optimal number of calls to information recorded on the hard disk. \par
\vspace{12pt}
\centerline{\bf {Examples of images reconstruction in arbitrary sections}} \vspace{6pt}
A human knee-joint (20 $ \times  $ 10  $  \times  $ 10 cm in 20  $  \mu  $m steps) and a rat knee-joints (with a pixel size of 8  $  \times  $ 8  $  \mu  $m and a pitch of 8  $  \mu  $m) are considered as an examples of the section constructions. The samples were obtains in compliance with all legal norms. These sets could be easily identified by color of casting material. Green and red filling correspond to the rat and human knee-joints, respectively. \par
\begin{figure}[H]
\begin{center}
\includegraphics[width=4.79in,height=1.24in]{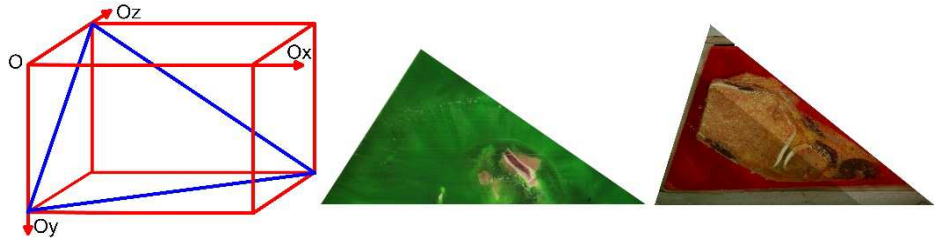}
\caption{Triangular sections}
\end{center}
\label{fig3}
\end{figure}
\begin{figure}[H]
\begin{center}
\includegraphics[width=4.79in,height=1.26in]{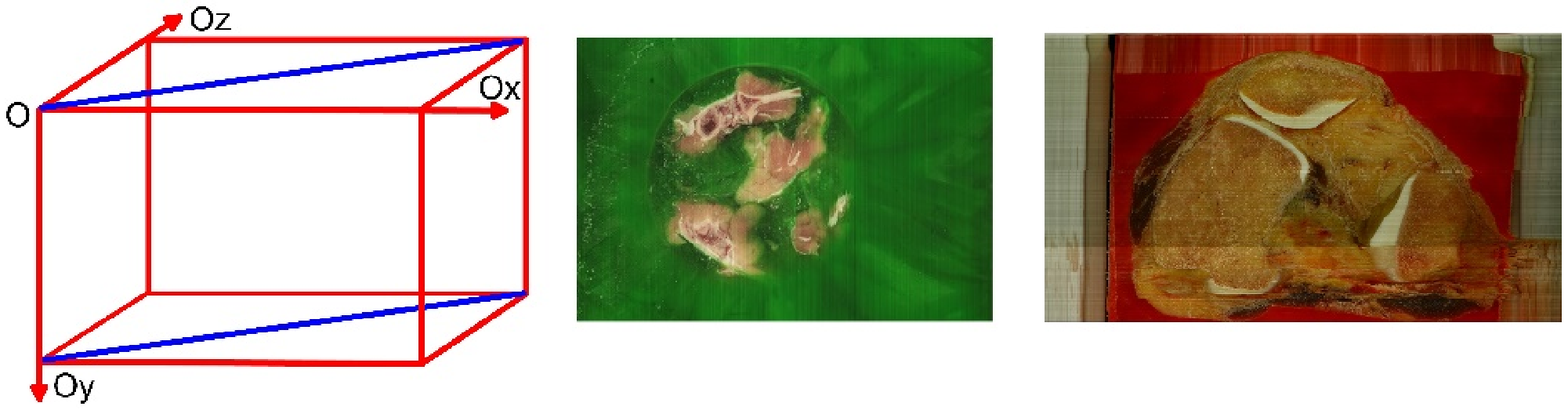}
\caption{Cross-sections with 4 angles}
\end{center}
\label{fig4}
\end{figure}
\begin{figure}[H]
\begin{center}
\includegraphics[width=4.8in,height=1.26in]{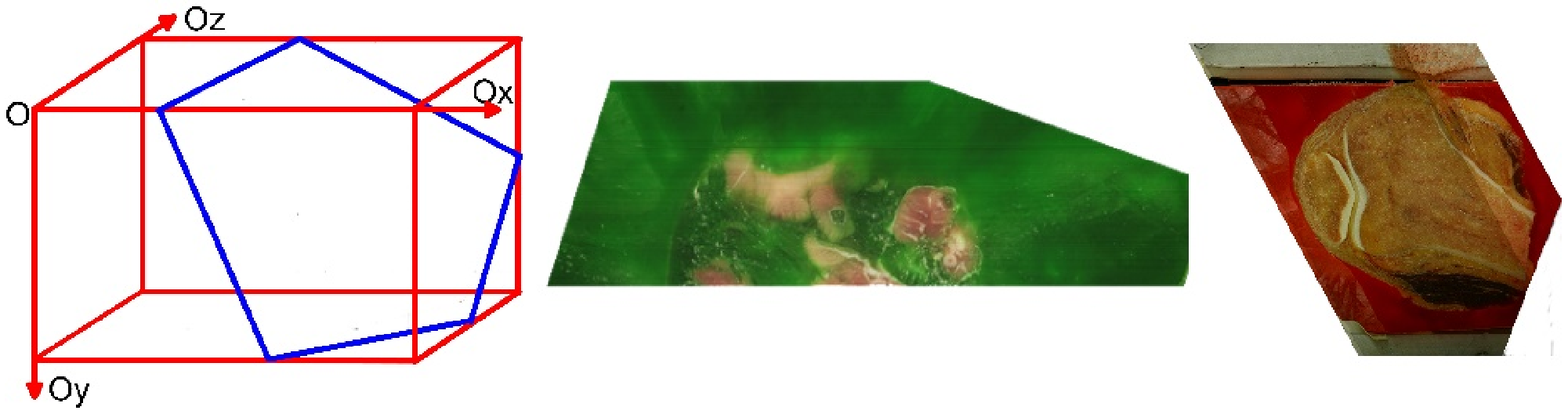}
\caption{Cross-sections with 5 angles}
\end{center}
\label{fig5}
\end{figure}
\begin{figure}[H]
\begin{center}
\includegraphics[width=4.8in,height=1.29in]{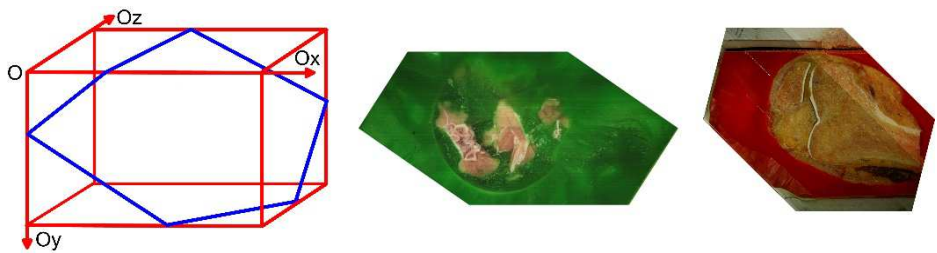}
\caption{Cross-sections with 6 angles}
\end{center}
\label{fig6}
\end{figure}

Images of different shapes of sections are considered below. If the section does not have the shape of a rectangle, then the image is supplemented with a black background (no restriction on the background color) (figures \ref{fig3}{3}, \ref{fig4}{4}, \ref{fig5}{5} and \ref{fig6}{6}). \par
\vspace{12pt}

\newpage
\centerline{\bf {Information system for modeling}} \vspace{6pt}

\begin{center}


\begin{figure}[H]
\begin{center}
\includegraphics[width=4.8in,height=3.06in]{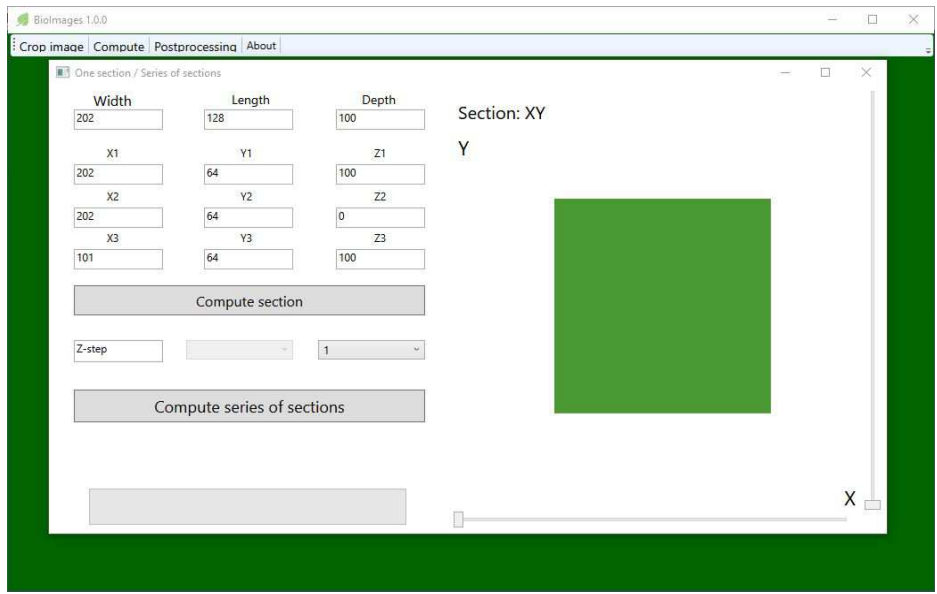}
\caption{Setting of an arbitrary slice}
\end{center}
\label{fig7}
\end{figure}


\end{center}

The developed software package includes two main parts: programs with a graphical user interface (GUI) and calculation modules (DLL) for modeling user-defined slices. Microsoft Visual Studio 2012 comprising C $  \#  $ programming language with .NET framework 4.5 has been utilized as the software development environment. The programming interface is refined by API WPF. User selects a folder with input data, sets the geometric parameters of the sample (length, width, height and step  $  \Delta  $\textit{z} as seen from Fig. \ref{fig7}{7}) and sections (three points along which the plane will be built). The file names are determined by the section number in the sequence. \par
\noindent
\vspace{12pt}
\noindent
\par After the previously described actions with the help of calculation modules, the program creates an object of the Cuboid class (with the specified length, width and depth) and an object of the Plane class describing the frontal face of the cuboid (the normal vector is co-directed with the OZ axis and applied to the origin --- (0, 0, 0)). \par
\begin{center}


\begin{figure}[H]
\begin{center}
\includegraphics[width=4.8in,height=1.91in]{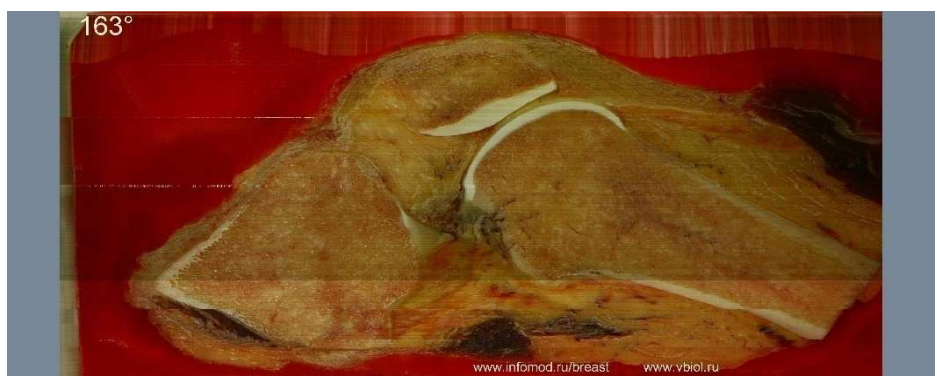}
\caption{Postprocessed image}
\end{center}
\label{fig8}
\end{figure}


\end{center}

 \hspace*{0.16in} Beside the one slide calculation, series of slices can be constructed (for example, series of images with a given step in the range 1-6  $  ^{\circ}  $). In the case if images size vary, they should be enlarged to the same size. In order to do this all the resulting slices are analyzed and the image with the largest width is selected. The remaining images are enlarged to chosen size and supplemented with a gray-blue background (Fig. \ref{fig8}{8}
). The resulting images are marked with angle values. Using the images movie can be compiled (for example, for XY- and XZ- sections (along the Z- and Y- axes, respectively)). \par
\noindent
\vspace{12pt}
\noindent
\centerline{\bf {Conclusions}}
\noindent
\par A software product has been developed allowing work with images of sections obtained by the method of high-precision grinding. The method of obtaining slices and the algorithm for their subsequent processing completely exclude spatial deformation and may be characterized by minimal color distortion. \par
To test the software complex and demonstrate the results of section construction, we used samples of human and rat knee-joints. The virtual model development of a human knee-joint may have practical application in new technologies of restoration of articular surfaces at joint injuries [\ref{b17}{17}, \ref{b18}{18}]. In addition, it may be applied for making of virtual simulators based on the principles of augmented reality or for creation of 3D-printer models. This method can be used for aims of medicine, veterinary medicine, zootechny and related areas where MRI or histological reconstructions do not provide a full-fledged three-dimensional view [\ref{b3}{3}, \ref{b4}{4}, \ref{b10}{10}]. \par
We can use this method to construct vector 3D models of samples that are necessary for theoretical modeling physical processes in highly heterogeneous biological tissues [\ref{b18}{18}, \ref{b21}{21}]. Such 3D models allow to take into account the fine structure of real tissues and changes in their physical parameters. The joint use of image processing techniques and numerical modeling methods such as computational fluid dynamics or heat transfer is a modern trend in the development of medicine and applied biology [\ref{b13}{13}]. \par
\textbf{Acknowledgments.} KAV and AAS are thankful to the Ministry of Education and Science of the Russian Federation (project No. 2.852.2017/4.6). NVV thanks the RFBR grant and Volgograd Region Administration (No. 15-47-02642). \par
\vspace{6pt}
\textbf{References} \par \par
\noindent
\begin{enumerate}
{\label{b1}\item\fontsize{9pt}{9pt}\selectfont  Azinfar, L., Ravanfar, M., Wang, Y., Zhang, K., Duan, D., Yao, G.: High resolution imaging
of the fibrous microstructure in bovine common carotid artery using optical polarization
tractography. J. Biophotonics 10, 231–241 (2017). doi: 10.1002/jbio.201500229} \par
{\label{b2}\item\fontsize{9pt}{9pt}\selectfont Bobroff, V., Chen, H.-H., Delugin, M., Javerzat, S., Petibois, C.: Quantitative IR microscopy
and spectromics open the way to 3D digital pathology. J. Biophotonics 10, 598–606 (2017).
doi: 10.1002/jbio.201600051} \par
{\label{b3}\item\fontsize{9pt}{9pt}\selectfont Brazina, D., Fojtik, R., Rombova, Z.: 3D visualization in teaching anatomy. Procedia Soc.
Behav. Sci. 143, 367–371 (2014).  doi: 10.1016/j.sbspro.2014.07.496} \par
{\label{b4}\item\fontsize{9pt}{9pt}\selectfont Candemir, S., Jaeger, S., Antani, S., Bagci, U., Folio, L.R., Xu, Z., Thoma, G.: Atlas-based
rib-bone detection in chest X-rays. Comput. Med. Imaging Graph. 51, 32–39 (2016).  doi: 10.1016/j.compmedimag.2016.04.002} \par
{\label{b5}\item\fontsize{9pt}{9pt}\selectfont Cerveri, P., Manzotti, A., Confalonieri, N., Baroni, G.: Automating the design of resection
guides specific to patient anatomy in knee replacement surgery by enhanced 3D curvature
and surface modeling of distal femur shape models. Comput. Med. Imaging Graph. 38(8),
664–674 (2014). doi: 10.1016/j.compmedimag.2014.09.001} \par
{\label{b6}\item\fontsize{9pt}{9pt}\selectfont Chen, Y., Wang, Z., Li, L., Wan, X., Sun, F., Zhang, F.: A fully automatic geometric parameters determining method for electron tomography. In: Cai, Z., Daescu, O., Li, M. (eds.)ISBRA 2017. LNCS, vol. 10330, pp. 385–389. Springer, Cham (2017).  doi: 10.1007/978-3-319-59575-7 $  \_  $39} \par
{\label{b7}\item\fontsize{9pt}{9pt}\selectfont Chiorean, L.D., Szasz, T., Vaida, M.F., Voina, A.: 3D reconstruction and volume computing
in medical imaging. Acta Technica Napocensis 52(3), 18–24 (2011).} \par
{\label{b8}\item\fontsize{9pt}{9pt}\selectfont Cuijpers V.M.J.I., Walboomers X.F., Jansen J.A.: Three-Dimensional Localization of Implanted Biomaterials in Anatomical and Histological Specimens Using Combined X-Ray Computed Tomography and Three-Dimensional Surface Reconstruction: A Technical Note. Tissue Eng. Part C Methods 16, 63-69 (2010). doi: 10.1089/ten.TEC.2008.0604} \par
{\label{b9}\item\fontsize{9pt}{9pt}\selectfont  Ermilov, S.A., Su, R., Conjusteau, A., Anis, F., Nadvoretskiy, V., Anastasio, M.A., Oraevsky,
A.A.: Three-dimensional optoacoustic and laser-induced ultrasound tomography system for
preclinical research in mice: design and phantom validation. Ultrason. Imaging 38, 77–95
(2016). doi: 10.1177/0161734615591163} \par
{\label{b10}\item\fontsize{9pt}{9pt}\selectfont Ha, J.F., Morrison, R.J., Green, G.E., Zopf, D.A.: Computer-aided design and 3-dimensional
printing for costal cartilage simulation of airway graft carving. Otolaryngol. Head Neck Surg.1–4 (2017).  doi: 10.1177/0194599817697048} \par
{\label{b11}\item\fontsize{9pt}{9pt}\selectfont  Hanney, M.B., Hillel, P.G., Scott, A.D., Lorenz, E.: Half-body single photon emission
computed tomography with resolution recovery for the evaluation of metastatic bone disease: implementation into routine clinical service. Nuclear Med. Commun. 38, 623–628 (2017). doi: 10.1097/MNM.0000000000000686} \par
{\label{b12}\item\fontsize{9pt}{9pt}\selectfont  Ioakemidou, F., Ericson, F., Spuhler, J., Olwal, A., Forsslund, J., Jansson, J., Pysander, E.-L.S.,
Hoffman, J.: Gestural 3D interaction with a beating heart: simulation, visualization and
interaction. In: Proceedings of SIGRAD 2011, KTH, Stockholm, pp. 93–97 (2011).} \par
{\label{b13}\item\fontsize{9pt}{9pt}\selectfont Ko, Z.Y.G., Mehta, K., Jamil, M., Yap, C.H., Chen, N.: A method to study the hemodynamics
of chicken embryo’s aortic arches using optical coherence tomography. J. Biophotonics 10,
353–359 (2017).  doi: 10.1002/jbio.201600119} \par
{\label{b14}\item\fontsize{9pt}{9pt}\selectfont  Lee, R.C., Darling, C.L., Staninec, M., Ragadio, A., Fried, D.: Activity assessment of root
caries lesions with thermal and near-IR imaging methods. J. Biophotonics 10, 433–445
(2017). doi: 10.1002/jbio.201500333} \par
{\label{b15}\item\fontsize{9pt}{9pt}\selectfont  Mohammed, I.M., Tatineni, J., Cadd, B., Gibson, I.: Advanced auricular prosthesis
development by 3D modelling and multi-material printing. In: Proceedings of the
International Conference on Design and Technology. DesTech Conference, Geelong, pp. 37–
43 (2017). doi: 10.18502/keg.v2i2.593} \par
{\label{b16}\item\fontsize{9pt}{9pt}\selectfont Murino, L., Granata, D., Carfora, M.F., Selvan, S.E., Alfano, B., Amato, U., La-robina, M.: Evaluation of supervised methods for the classification of major tissues and sub-cortical structures in multispectral brain magnetic resonance images. Comput. Med. Imaging Graph.38(5), 337–347 (2014).  doi: 10.1016/j.compmedimag.2014.03.003} \par
{\label{b17}\item\fontsize{9pt}{9pt}\selectfont  Novochadov, V.V., Khoperskov, A.V., Terpilovskiy, A.A., Malanin, D.A., Tiras, K.P., Kovalev, M.E., Astakhov, A.S.: Virtual full-color three-dimensional reconstruction of human knee joint by the digitization of serial layer-by-layer grinding. In: Mathematical Biology and Bioinformatics. Reports of the VI International Conference, Puschino, pp. 76–78 (2016).} \par
{\label{b18}\item\fontsize{9pt}{9pt}\selectfont  Novochadov, V.V., Shiroky, A.A., Khoperskov, A.V., Losev, A.G.: Comparative modeling
the thermal transfer in tissues with volume pathological focuses and tissue engineering
constructs: a pilot study. Eur. J. Mol. Biotechnol. 14, 125–138 (2016). doi: 10.13187/ejmb.2016.14.125} \par
{\label{b19}\item\fontsize{9pt}{9pt}\selectfont  Novochadov, V.V., Terpilovsky, A.A., Shirokiy, A.A., Tiras, K.P., Klimenko, A.S.,
Klimenko, S.V.: Visual analytics based on recoding input color information in 3D-
reconstructions of human bones and joint. In: C-IoT-VRTerro 2016, pp. 257–260. Institute
of Physical and Technical Informatics, Protvino (2016).} \par
{\label{b20}\item\fontsize{9pt}{9pt}\selectfont Papantoniou, I., Sonnaert, M., Geris, L., Luyten, F.P., Schrooten, J., Kerck-hofs, G.: Three-
dimensional characterization of tissue-engineered constructs by contrast-enhanced nanofocus computed tomography. Tissue Eng. Part C Methods 20, 177–187 (2014). doi: 10.1089/ten.TEC.2013.0041} \par
{\label{b21}\item\fontsize{9pt}{9pt}\selectfont Polyakov, M.V., Khoperskov, A.V.: Mathematical modeling of radiation fields in biological tissues: the definition of the brightness temperature for the diagnosis. Sci. J. VolSU Math. Phys. 5(36), 73–84 (2016). doi: 10.15688/jvolsu1.2016.5.7} \par
{\label{b22}\item\fontsize{9pt}{9pt}\selectfont T Terpilovskij, A.A., Kuz’min, A.L., Lukashkina, R.A.: Method for creating a virtual model of a biological object and a device for its implementation. Patent of the Russian Federation. Invention No. 2418316, 10 May 2011. Bull. 13.} \par
{\label{b23}\item\fontsize{9pt}{9pt}\selectfont Terpilovskiy, A.A., Tiras, K.P., Khoperskov, A.V., Novochadov, V.V.: The possibilities of
full-color three-dimensional reconstruction of biological objects by the method of layer-by-layer overlapping: knee joint of a rat. Sci. J. Volgograd State Univ. Nat. Sci. 4, 6–14 (2015). doi: 10.15688/jvolsu11.2015.4.1} \par
{\label{b24}\item\fontsize{9pt}{9pt}\selectfont Turlapov, V.E., Gavrilov, N.I.: 3D scientific visualization and geometric modeling in digital
biomedicine. Sci. Vis. 7(4), 27–43 (2015).} \par
{\label{b25}\item\fontsize{9pt}{9pt}\selectfont Uma Vetri Selvi, G., Nadarajan, R.: A rapid compression technique for 4-D functional MRI images using data rearrangement and modified binary array techniques. Australas. Phys. Eng.Sci. Med. 38, 731–742 (2015). doi: 10.1007/s13246-015-0385-y} \par
{\label{b26}\item\fontsize{9pt}{9pt}\selectfont Weber, L., Langer, M., Tavella, S., Ruggiu, A., Peyrin, F.: Quantitative evaluation of
regularized phase retrieval algorithms on bone scaffolds seeded with bone cells. Phys. Med.Biol. 61, 215–231 (2016). doi: 10.1088/0031-9155/61/9/N215} \par
{\label{b27}\item\fontsize{9pt}{9pt}\selectfont Xu, X., Chen, X., Li, F., Zheng, X., Wang, Q., Sun, G., Zhang, J., Xu, B.: Effectiveness of
endoscopic surgery for supratentorial hypertensive intracerebral hemorrhage: a comparison
with craniotomy. J. Neurosurg. 1–7 (2017).  doi: 10.3171/2016.10.JNS161589} \par
\vspace{12pt}
\end{enumerate}
\end{document}